\documentclass[runningheads]{llncs}
\usepackage{hyperref} %hidelinks, 
\hypersetup{hidelinks=false,urlcolor=blue}
\usepackage{graphicx}
\usepackage{booktabs}%
\usepackage{subcaption}
\usepackage{xurl}
\usepackage{array}
\newcolumntype{C}[1]{>{\centering\arraybackslash}p{#1}}

\begin{document}
\title{SoftCTM: Cell detection by soft instance segmentation and consideration of cell-tissue interaction}
\titlerunning{Cell detection by soft instance segmentation and cell-tissue interaction}
\author{Lydia A. Schoenpflug\inst{1,^*}\orcidID{0009-0009-6703-9368} \and
Viktor H. Koelzer\inst{1}\orcidID{0000-0001-9206-4885}}
\authorrunning{L. A. Schoenpflug and V. H. Koelzer}
\institute{\inst{1}Department of Pathology and Molecular Pathology, University Hospital and University of Zürich, Schmelzbergstrasse 12, 8091 Zürich, Switzerland *Corresponding author, \email{lydia.schoenpflug@usz.ch}}
\maketitle              % typeset the header of the contribution

\begin{abstract}
Detecting and classifying cells in histopathology H\&E stained whole-slide images is a core task in computational pathology, as it provides valuable insight into the tumor microenvironment. In this work we investigate the impact of ground truth formats on the models performance. Additionally, cell-tissue interactions are considered by providing tissue segmentation predictions as input to the cell detection model. We find that a “soft”, probability-map instance segmentation ground truth leads to best model performance. Combined with cell-tissue interaction and test-time augmentation our Soft Cell-Tissue-Model (SoftCTM) achieves 0.7172 mean F1-Score on the Overlapped Cell On Tissue (OCELOT) test set, achieving the third best overall score in the OCELOT 2023 Challenge. The source code for our approach is made publicly available at \url{https://github.com/lely475/ocelot23algo}.
\end{abstract}

\section{Introduction}\label{Introduction}
Cell detection and classification is a sub-task of Computational Pathology, which can be achieved through deep learning as shown in \cite{Sun2021,Graham2019,Naylor2019,Liu2019,Zeng2019,Bancher2021}. Jeongun Ryu and colleagues demonstrate in \cite{Ryu2023} that cell detection can benefit from considering cell-tissue relationships. They furthermore introduce the OCELOT dataset, which consists of 667 pairs of high resolution patches for cell detection in combination with lower resolution tissue patches, showing additional tissue context around the cell patch. The OCELOT 2023: Cell Detection from Cell-Tissue Interaction Challenge \cite{grandchallengeOCELOT2023} motivates the development of cell detection algorithms that take the surrounding tissue context into account, based on the OCELOT dataset. In this paper, we present an approach to the OCELOT 2023 Challenge. %  (Figure \ref{fig:approach}). 
First, we investigate the impact of different ground truth formats on the model performance. Second, we utilize the tissue segmentation annotation, by training a second model for tissue segmentation and providing its predictions as input to the cell detection model. Third, we utilize test-time augmentation (TTA), to further improve the model. Our final approach achieves the third best mean F1 Score of 0.7172 on the OCELOT test set.

\section{Related Works}\label{Related Works}
Deep learning approaches to cell detection in histopathology can be categorized into (1) Semantic segmentation-based approaches paired with instance extraction by either (a) postprocessing steps, such as a watershed transform \cite{Graham2019,Naylor2019}, local maxima extraction \cite{Ryu2023} or morphological operations \cite{Zeng2019}, or (b) an object detection network \cite{Bancher2021,Liu2019}, and (2) pure object-detection approaches \cite{Sun2021}. 
All approaches except \cite{Ryu2023} are trained on cell annotations only. In contrast, \cite{Ryu2023} motivates the consideration of tissue context for cell detection, demonstrating improved generalization for the OCELOT dataset.
As the OCELOT dataset provides only cell centroid annotations, \cite{Ryu2023} translates them into a cell segmentation map by assigning pixels within a fixed radius of the nuclei centroid to the cells class. This potentially limits the model training, as only consideration of pixels around the nuclei centroid is rewarded. In contrast, \cite{Graham2019,Naylor2019} draw on full instance segmentation ground truths, enabling the consideration of all nuclei pixels. Furthermore, \cite{Naylor2019} translates the ground truth into a cell probability map instead of class labels to better reflect the cells blurry boundaries and enable a smoother prediction.
This motivated us to extend \cite{Ryu2023}, by enriching the ground truth formats from point annotations to instance segmentation maps and then investigating different ground truth formats for semantic segmentation.

\section{Methods}\label{Methods}
In this section we describe the dataset and configuration for training a tumor segmentation model, as well as training a cell detection model on three different ground truth formats. Our proposed final workflow is a combined cell-tissue model (CTM) as described in Section \ref{Combined Cell-Tissue Model}. 

\subsection{Dataset}\label{Dataset Preparation and Partitioning}
The OCELOT training dataset \cite{Ryu2023} consists of 400 pairs of cell and tissue patches of size $1024 \times 1024$, with a magnification of 50x and 12.5x respectively.
Cell annotations are provided as nuclei centroid coordinates ("cell point annotation") for the classes tumor and background cells. %, with a ratio of 64.6\% and 35.4\%. 
Tissue annotations are provided as pixel-wise segmentation masks for the classes cancer area, background and unknown. %, with a ratio of 39.8\%, 55.8\% and 4.3\%.
We split the dataset into an internal training and validation set on a 80:20 split with stratified cell and tissue annotation and organ distribution (training set: 320 patch pairs from 138 WSIs, validation set: 80 patch pairs from 35 WSIs). The internal validation set is utilized for hyperparameter optimization. All further experiments are validated on the OCELOT validation set \cite{Ryu2023}.

\subsection{Model Architecture}
\label{subsubsection: Model architecture}
\begin{figure}[htbp]
    \centering
    \includegraphics[width=0.92\textwidth]{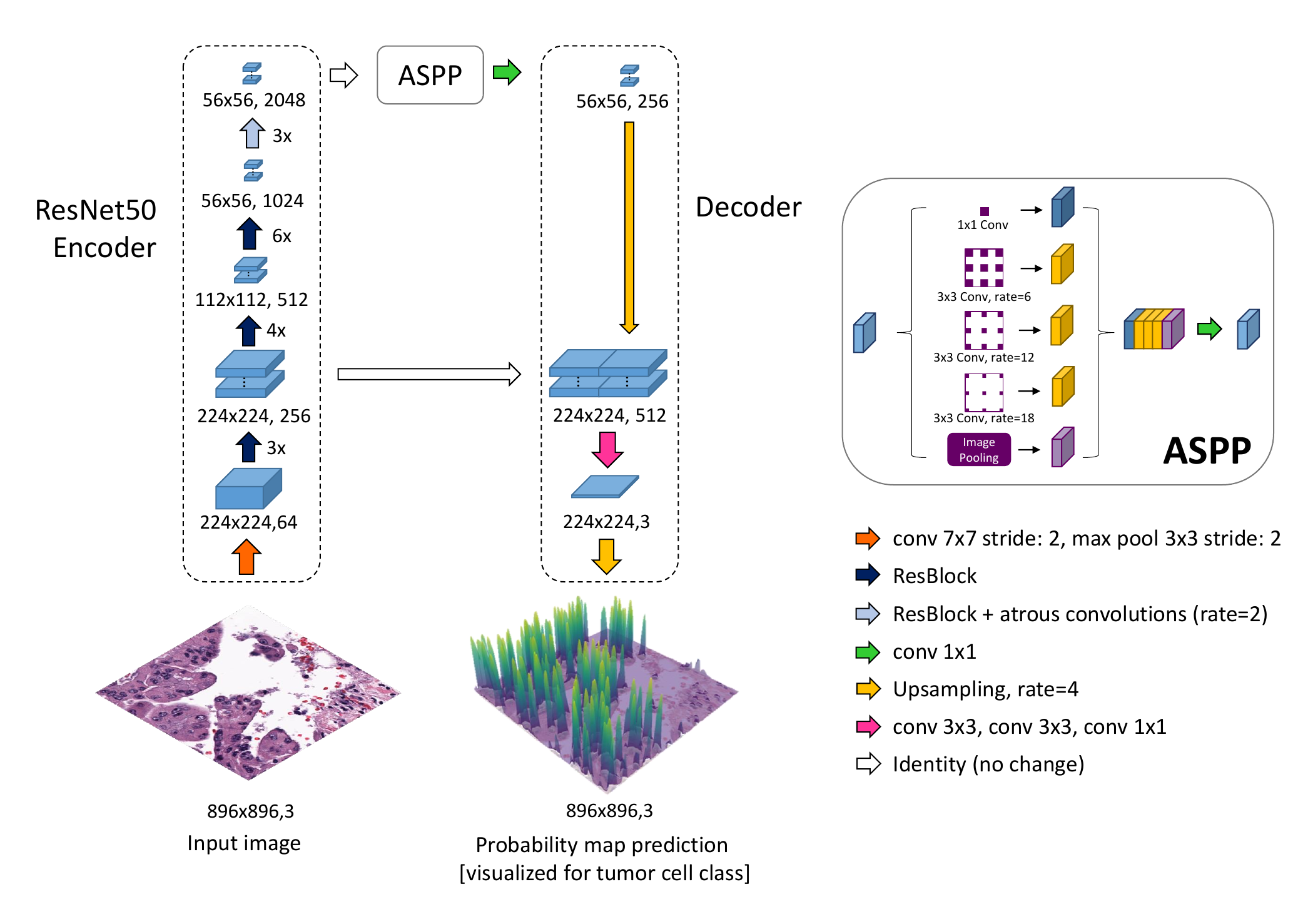}
    \caption{Model architecture: DeepLabv3+ with ResNet50 Encoder, an Atrous Spatial Pyramid Pooling (ASPP) block and a Decoder with 2 upsampling and 3 convolutional blocks.} %The input to the model is the image, or the image combined with the tissue segmentation prediction for the }
    \label{fig:arch}
\end{figure}

Building on \cite{Ryu2023} we utilize a DeepLabv3+ segmentation model \cite{DeepLabv3plus2018} with a ResNet50 \cite{ResNet2016} encoder pretrained on ImageNet\footnote{Pytorch default pretrained weights: \url{https://pytorch.org/vision/stable/models/generated/torchvision.models.resnet50.html\#torchvision.models.ResNet50_Weights}, last accessed 24.11.2023} as shown in Figure \ref{fig:arch}. 
The last ResNet50 convolutional block utilizes atrous convolutions with a rate of 2, to enable downsampling while preserving the input feature dimension. This is  followed by an Atrous Spatial Pyramid Pooling (ASPP) block \cite{DeepLab} and a decoder. The ASPP block extracts high-level-features at different downsampling rates to account for differing object scales. The decoder consists of two upsampling steps, connected by three convolutional layers which combine low-level features from the second ResNet50 layer with high-level features from the upsampled ASPP output. We utilize the same model architecture for cell detection and tissue segmentation.

\subsection{Tissue Segmentation Training}
Training hyperparameters for the tissue segmentation model were based on \cite{schoenpflug2023multitask}. The model was trained for 100 epochs on the internal training set to minimize a cross entropy loss with stochastic gradient descent (initial learning rate: 0.2, Nesterov momentum: 0.9, weight decay: $5\cdot 10^{-6}$, exponential learning rate decay: $\gamma=0.97$) and a batch size of 8. The best model was selected based on the internal validation set. Training samples were oversampled to achieve a balanced amount of background and cancer pixels. The input images were augmented by re-scaling in the range $\pm 10\%$, random crop to $896\times896$ pixels, flip, rotation by 90°, 180° or 270° and a channel-wise brightness and contrast variation by $\pm 20\%$. Each augmentation is applied with a probability of 70\%.

\subsection{Cell Detection Ground Truth Formats}
\begin{figure}[htbp]
    \centering
    \begin{subfigure}[t]{0.48\textwidth}
        \centering
        \includegraphics[width=0.92\textwidth]{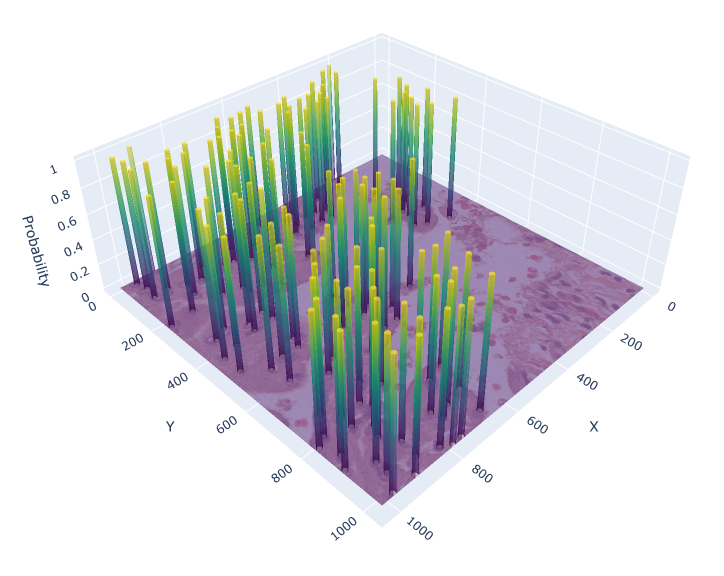}
        \caption{}
        \label{fig:subfig1}
    \end{subfigure}
    \hfill
    \begin{subfigure}[t]{0.48\textwidth}
        \centering
        \includegraphics[width=0.92\textwidth]{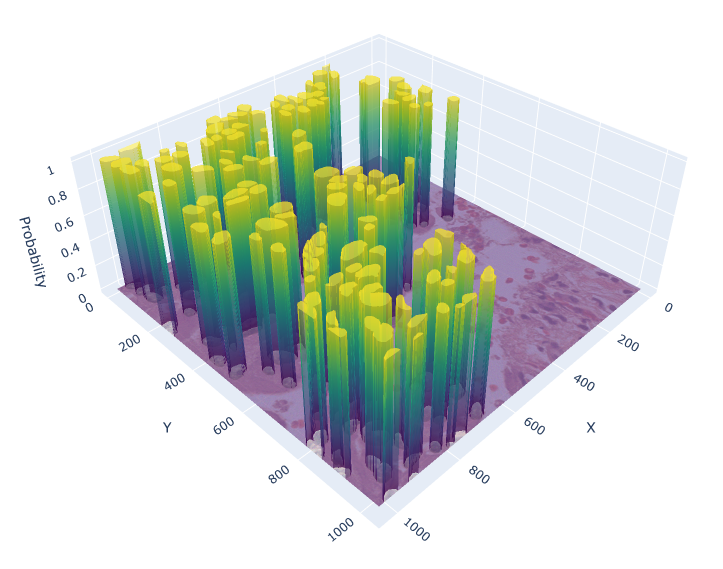}
        \caption{}
        \label{fig:subfig2}
    \end{subfigure}

    \begin{subfigure}[t]{0.48\textwidth}
        \centering
        \includegraphics[width=0.92\textwidth]{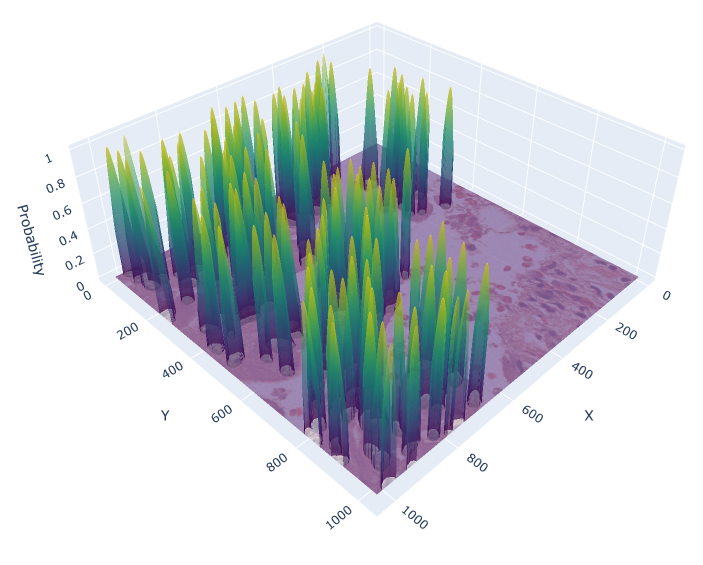}
        \caption{}
        \label{fig:subfig3}
    \end{subfigure}
    \hfill
    \begin{subfigure}[t]{0.48\textwidth}
        \centering
        \includegraphics[width=0.6\textwidth]{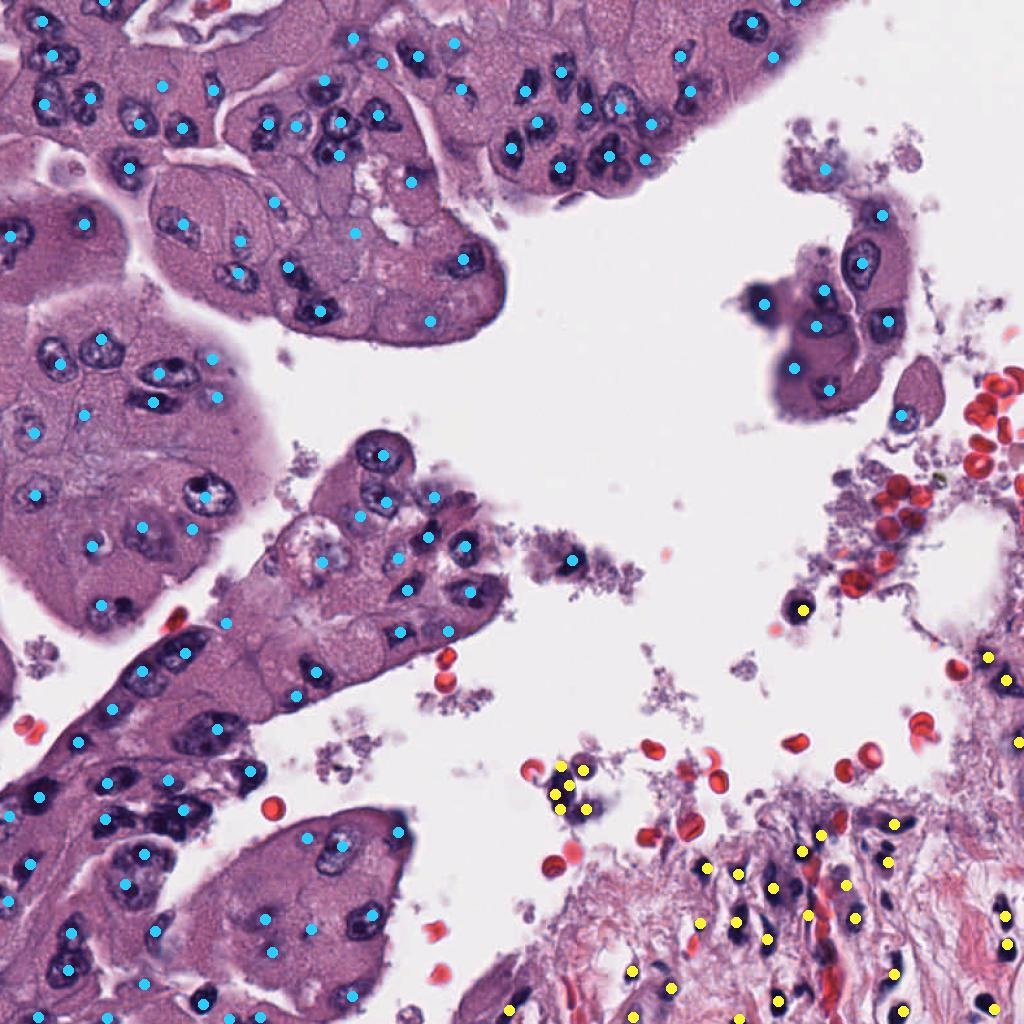}
        \caption{}
        \label{fig:subfig4}
    \end{subfigure}
    \caption{Example of different ground truth formats, visualization of the tumor cell class probability map: (a) Circle, (b) Hard IS, (c) Soft IS, (d) Original image with cell point annotation (blue: tumor cells, yellow: background cells)}
    \label{fig:main}
\end{figure}

The sparse cell point annotations require translation into segmentation annotations.
We investigate the following ground truth formats:
\begin{enumerate}
    \item \textbf{Circle: }Identical to \cite{Ryu2023}, all pixels in a circle, centered on the cell coordinates with radius of $1.4\mu m$ (r=7 pixels at 0.2 microns-per-pixel (mpp) magnification), are assigned to the cells class id (Figure \ref{fig:subfig1}).
    \item \textbf{Hard instance segmentation (Hard IS):} Inspired by \cite{Jahanifar2021} instance segmentation masks are derived from the image and centroid coordinates by applying NuClick\footnote{Publicly available at \url{https://github.com/navidstuv/NuClick}, last accessed 24.11.2023}, a CNN-based segmentation model \cite{AlemiKoohbanani2020} that utilizes the centroid coordinates for nucleus instance segmentation (more details in Appendix \ref{appdx: NuClick}). All pixels belonging to a nucleus instance are assigned the cells class label (Figure \ref{fig:subfig2}). If the NuClick model was not able to segment a cells nucleus, we revert to the circle ground truth format. % and place a circle with radius $1.4\mu m$ instead.
    \item \textbf{Soft instance segmentation (Soft IS):} Motivated by \cite{Naylor2019}, we place a Gaussian with $\sigma=3\mu m$ (15 pixels at 0.2 mpp, different $\sigma$ values are investigated in Appendix \ref{appdx:sigma}), centered at the centroid of each cell nuclei and set all background pixels not belonging to any NuClick cell instance to zero (Figure \ref{fig:subfig3}). This results in a probability map for each class, where the background probability map is derived as the inverse of the combined cell probability maps $y_{bg} = 1 - \sum_c y_{c}$.
\end{enumerate}

\subsection{Cell Detection Training} Model hyperparameters such as learning rate, optimizer, architecture and loss function were selected based on performance on the internal validation set. For the learning rate we considered values in the range [$5\cdot 10^{-5}$, $2\cdot 10^{-3}$]. The model was trained on the OCELOT training set in a k-fold manner with
fixed hyperparameters (k=5), resulting in an 80:20 training to validation split for each iteration. The five trained models were combined by using the averaged sum of their predictions.
The cell detection models were trained for 150 epochs with a weighted Adam optimizer (learning rate: $8\cdot 10^{-4}$) and a batch size of 32. The learning objective is minimizing a Dice loss \cite{DiceLoss} for the circle and hard IS ground truth format:
\begin{equation}
    Generalized~Dice~Loss = 1 - 2 \cdot \frac{\sum_{c} w_c \cdot \sum_{i=1}^N  y_{i,c} \cdot \hat{y}_{i,c}}{\sum_{c} w_c \cdot \sum_{i=1}^N y_{i,c} + \hat{y}_{i,c}} ,~~~w_c = \frac{1}{\sum_{i} y_{i,c}}
\end{equation}
Where $y_{i,c}$ is the segmentation ground truth and $\hat{y}_{i,c}$ is the segmentation prediction for pixel i and class $c$, with a weighting of $w_c$ for each class.
The soft IS format poses a pixel-wise regression problem, for this reason we utilize a weighted mean square error (MSE) loss :
\begin{equation}
    Weighted~MSE~Loss =\sum_{c} w_c \cdot \frac{1}{N} \sum_{i=1}^N (y_{i,c} - \hat{y}_{i,c})^2,~~~w_c = \frac{\sum_{i} y_i}{\sum_{i} y_{i,c}}
\end{equation}
Training samples were over-sampled based on the presence of background and tumor cells in each sample, to achieve a balanced number of tumor and background cell instances. The same augmentation methods as for the tumor segmentation model were applied.

\subsection{Cell Detection Postprocessing}
\begin{figure}[htbp]
    \centering
    \includegraphics[width=0.92\textwidth]{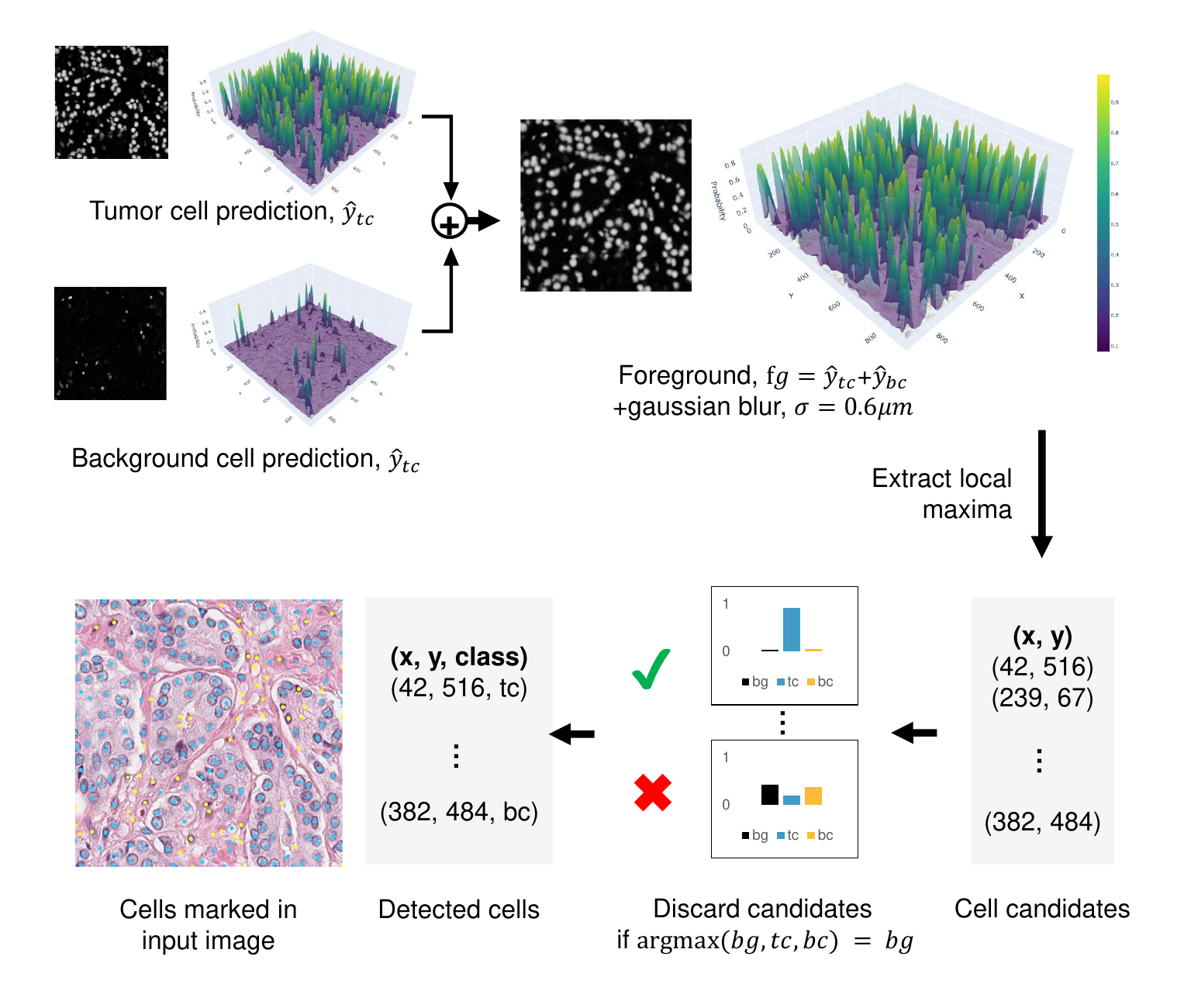}
    \caption{Postprocessing for circle and soft IS ground truth format, tc: tumor cell class, bc: background cell class} %The input to the model is the image, or the image combined with the tissue segmentation prediction for the }
    \label{fig:postprocessing}
\end{figure}

For the circle and soft IS ground truth formats we extract cell detection candidates from the segmentation prediction by applying \textit{skimage.feature.peak\_local\_max} on the blurred foreground prediction (Figure \ref{fig:postprocessing}), where:
\begin{equation}
    foreground = \hat{y}_{tc} +\hat{y}_{bc} 
\end{equation}
$\hat{y}_{tc}$ and $\hat{y}_{bc} $ are the model predictions for the tumor cell class and background cell class respectively.
Only cell candidates with a larger probability for the tumor or background cell class, compared to the background class, are considered. The cell candidate class is assigned as the foreground class with the highest probability.
The hard IS requires a different approach, as cell instances are not trained to express a peak at the cell center and tend to overlap. For this reason, markers are extracted from the foreground prediction and then applied in a marker-controlled watershed segmentation to separate touching instance (more details in Appendix \ref{appx:nuclick-postprocessing}). The cell is assigned a class by majority vote of its pixel class predictions.

\subsection{Combined Cell-Tissue Model}\label{Combined Cell-Tissue Model}
\begin{figure}[htbp]
    \centering
    \includegraphics[width=\textwidth]{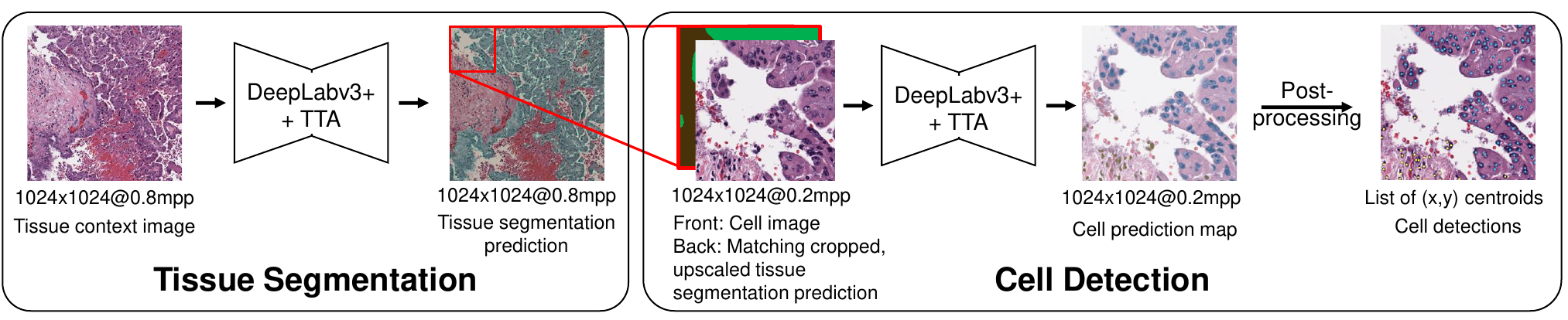}
    \caption{Combined cell-tissue model for cell detection: First, tissue segmentation is performed on a image showing surrounding tissue context of the cell patch. The tissue segmentation prediction is cropped and upsampled to match the cell patch, together forming the input to the cell detection model. Second, cell detection is performed resulting in a cell prediction map, which is postprocessed to extract class-wise cell centroid coordinates.}
    \label{fig:approach}
\end{figure}

We combine the cell detection and tissue segmentation models, by providing and re-training the cell detection model with both the cell patch and the cropped and upsampled tissue segmentation prediction as input (Figure \ref{fig:approach}). The cell detection model ground truth configuration is chosen based on the best performance on the OCELOT validation set. Additionally, we evaluated the effect of geometrical TTA for more robust predictions, consisting of all 8 possible rotation and flip combinations. TTA was applied for both models.

\section{Results}\label{results}
We evaluate the performance of the cell detection models trained on the three ground truth formats, as well as the performance of the CTM. Lastly, we investigate the effect of adding TTA.
\subsection{Main findings}
Table \ref{tab:mean-f1-across-sets} details the mean F1 score for the three ground truth formats. The best performing ground truth format is the soft IS across all sets. It is notable that the OCELOT test set shows a performance decrease from circle to hard IS ground truth. This is not the case for the internal validation and OCELOT validation set, but nevertheless highlights that while the hard IS ground truth increases the number of cell class pixels, this does not necessarily translate into a better model performance.
In contrast, utilizing the soft IS leads to a performance increase on all sets, possibly due to rewarding a local maxima, clearer cell boundaries and a simplified postprocessing.
As the soft IS showed highest performance among the three ground truth format models, we train the CTM with the soft IS ground truth for cell detection, further referred to as the SoftCTM.
The SoftCTM shows an increased mean F1 for the internal validation and OCELOT test set, but reduced mean F1 for the OCELOT validation set. 
However, combining the SoftCTM with test-time augmentation leads to the overall best score on the validation and test set. The need for TTA to improve performance on the validation set, when utilizing the SoftCTM, indicates that the models are to a certain extent sensitive to geometric variations. Yet, this is not observed on the internal validation and test set. 
\begin{table}[htb]
 \caption{Comparison of mean F1 scores for cell detection. The CTM was trained with the soft IS ground truth, as this showed the highest performance among the ground truth formats and is therefore denoted as SoftCTM. The highest score for the three ground truth formats is underlined, the overall highest score is marked in \textbf{bold}. We report the mean and std of the 5-Fold internal cross validation runs. }
  \centering
    \begin{tabular}{l C{3cm}C{2.5cm}C{2.5cm}}
    \midrule
    & 5-Fold  & OCELOT  & OCELOT \\
    &Internal validation& validation & test \\
     \midrule
    Circle        & .5647$\pm$.0262 & .6781 & .6599\\
   Hard IS & .6029$\pm$.0442 & .6826 & .6516\\
   \textbf{Soft IS} & \underline{.6494$\pm$.0302} & \underline{.6937} & \underline{.6777} \\
   \hline
   SoftCTM & .6842$\pm$.0238 & .6875 & .7090 \\
   SoftCTM + TTA & \textbf{.6950$\pm$.0245} & \textbf{.7046} & \textbf{.7172} \\
    \bottomrule
  \end{tabular}
  \label{tab:mean-f1-across-sets}
\end{table}
\begin{table}[htb]
 \caption{Comparison of mean F1 scores for tissue segmentation. We report the performance on the fixed internal validation set, as well as the OCELOT validation and test set. The highest score for each set is marked in \textbf{bold}.}
  \centering
    \begin{tabular}{l C{3cm}C{2.5cm}C{2.5cm}}
    \midrule
    & Internal & OCELOT  & OCELOT \\
    & validation & validation & test \\
     \midrule
    Tissue segmentation & .9131 & .8511 & .8927 \\
    Tissue segmentation + TTA  & \textbf{.9174} & \textbf{.8571} & \textbf{.8951} \\
    \bottomrule
  \end{tabular}
  \label{tab:mean-f1-tumor-sgm-across-sets}
\end{table}
\subsection{Organ-wise results}
Table \ref{tab:mean-f1-test-organ} provides insight into the per-organ mean F1 for cell detection and tissue segmentation on the OCELOT test set, for the validation set organ-wise performance is reported in Appendix \ref{appdx:organ-validation}. For kidney, endometrium, stomach and head-neck we observe the same tendencies as in the full set. In contrast, for prostate the three ground truth formats show only very little difference in performance, with the circle ground truth outperforming the others, while using the SoftCTM and TTA improves model performance. This is not the case for bladder, where the SoftCTM results in lower performance. We suspect this might be related to the less accurate tissue prediction for bladder, with approximately 0.84 F1-Score in contrast to $>0.9$ for the majority of organs (Table \ref{tab:mean-f1-test-organ-tissue-sgm}). However, prostate samples show an even lower tissue segmentation performance, yet utilizing the SoftCTM has a positive effect on the cell detection performance. We investigated this further by using the tissue segmentation ground truth as input to the SoftCTM\footnote{The tissue segmentation prediction was kept for all ground truth pixels of class Unknown and only replaced for the Background and Cancer Area pixels.}, we refer to this mode as the Tissue-label leaking model (TLLM). The largest improvement of utilizing the TLLM is observed for bladder samples with +10\% mean F1 score, which confirms that faulty tissue segmentation predictions lead to degraded cell detection performance. At the same time, we note that using the TLLM lead to a slight performance decrease compared to the SoftCTM for organs endometrium, stomach and head-neck, indicating that the tissue segmentation model appears to better capture the tissue composition than the ground truth for this subset.
\begin{table}[htb]
 \caption{Comparison of per-organ mean F1 scores on the OCELOT test set for cell detection. The CTM trained with the soft IS ground truth is denoted as SoftCTM. The tissue-label leaking model is denoted as TLLM. The highest score for the three ground truth formats is \underline{underlined}, the overall highest score is marked in \textbf{bold}.}
  \centering
  \begin{tabular}{lccccccc}
    \midrule
    & all & kidney & endometrium & bladder & prostate & stomach & head-neck \\
    & n=130 & n=41 & n=25 & n=26 & n=16 & n=12 & n=10 \\
     \midrule
    Circle        & .6599 & .6457 & .6791 & .6317 & \underline{.6276} & .6573 &  .6547\\
   Hard IS & .6516 & .6551 & .6486  & .6276 & .6228 & .6593 & .6624\\
   \textbf{Soft IS} & \underline{.6777} & \underline{.6608} & \underline{.7087}  & \underline{\textbf{.6515}} & .6220 & \underline{.6911} & \underline{.6966} \\
   \hline
    SoftCTM & .7090 & .7208 & .7527  & .6240 & .6416 & .7126 & .7457 \\
    SoftCTM + TTA & \textbf{.7172} & \textbf{.7323} & \textbf{.7560} & .6291  & \textbf{.6525} & \textbf{.7360} & \textbf{.7474} \\
   \hline
    TLLM & .7269 & .7409 & .7225  & .7259 & .6921 & .7092 & .7314 \\
    TLLM + TTA & .7315 & .7469 & .7210 & .7300  & .7063 & .7219 & .7338 \\
    \bottomrule
  \end{tabular}
  \label{tab:mean-f1-test-organ}
\end{table}
\begin{table}[htb]
 \caption{Comparison of per-organ mean F1 scores on the OCELOT test set for tissue segmentation. The tissue segmentation model is denoted as TSM. The highest score for each organ is marked in \textbf{bold}.}
  \centering
  \begin{tabular}{lccccccc}
    \midrule
    & all & kidney & endometrium & bladder & prostate & stomach & head-neck \\
    & n=130 & n=41 & n=25 & n=26 & n=16 & n=12 & n=10 \\
     \midrule
    TSM & .8927 & .9241 & .9062 & .8358  & .8119 & .9014  & \textbf{.8863} \\
    TSM + TTA & \textbf{.8951} & \textbf{.9282} & \textbf{.9088}  & \textbf{.8383} & \textbf{.8178} & \textbf{.9035} & .8774 \\
    \bottomrule
  \end{tabular}
  \label{tab:mean-f1-test-organ-tissue-sgm}
\end{table}
\section{Conclusion}\label{sec13}
Among the studied ground truth formats, extending the cell ground truth with the NuClick segmentation prediction and utilizing soft instance segmentation maps lead to the largest improvement in model generalization. By further combining it with tissue predictions and test-time augmentation we achieve 0.7172 mean F1 Score on the OCELOT test set. This supports the assumption that cell detection can benefit from considering the tissue context. As the OCELOT validation and test set originate from similar domains and are limited in size, future work could be conducted on evaluating the models usability on a larger cohort with respect to downstream tasks such as cell content estimation.


\bibliography{refs}
\bibliographystyle{splncs04}





% LNCS Homepage, \url{http://www.springer.com/lncs}. Last accessed 4
% \end{thebibliography}
\newpage
\appendix
\section*{Appendix}
We provide the following supplementary material:
\begin{itemize}
    \item Description of segmentation ground truth generation with NuClick
   \item Postprocessing steps when using the hard segmentation ground truth format
    \item An Ablation study considering different $\sigma$ values for the soft IS ground truth
    \item Organ-wise performance on OCELOT validation set
\end{itemize}

\section{Segmentation ground truth generation with NuClick}
\label{appdx: NuClick}
\begin{figure}[htbp]
    \centering
    \begin{subfigure}[t]{0.3\textwidth}
        \centering
        \includegraphics[width=0.9\textwidth]{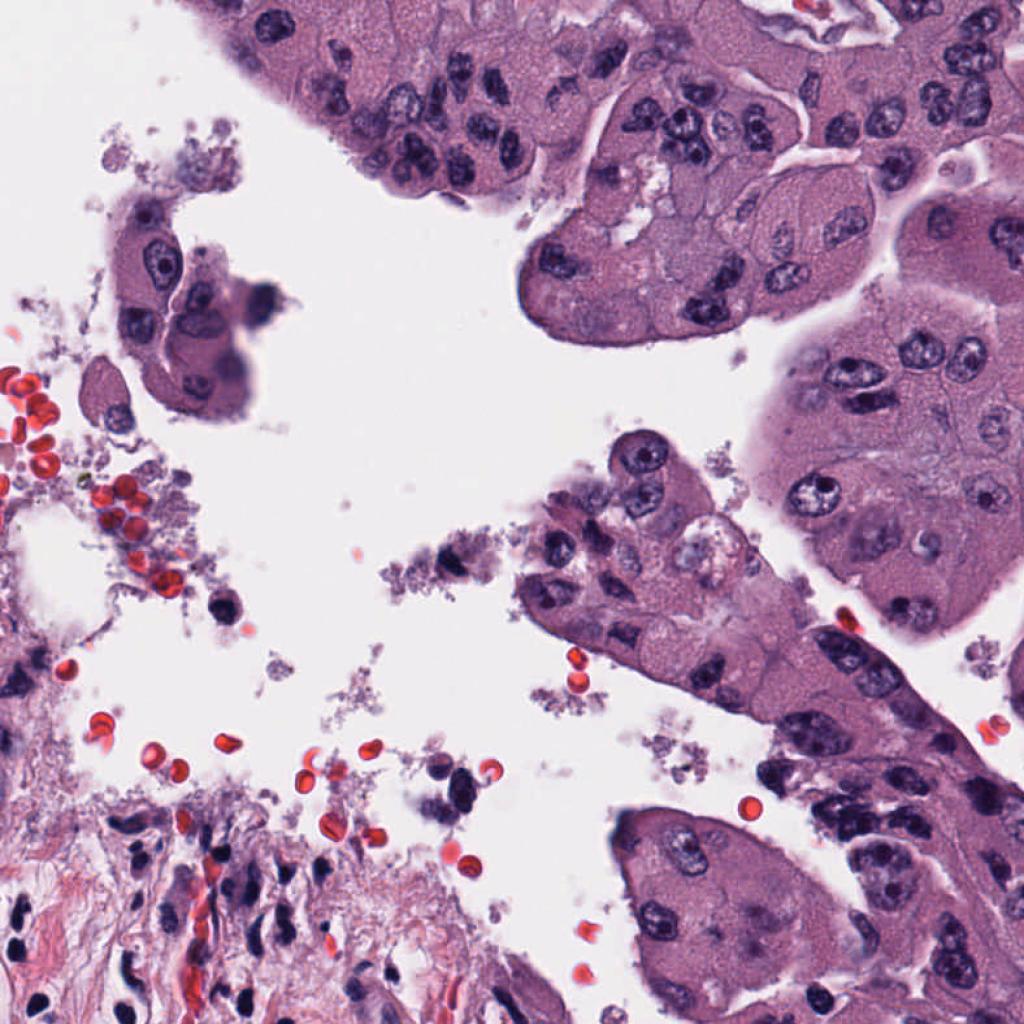}
        \caption{}
    \end{subfigure}
    \begin{subfigure}[t]{0.3\textwidth}
        \centering
        \includegraphics[width=0.9\textwidth]{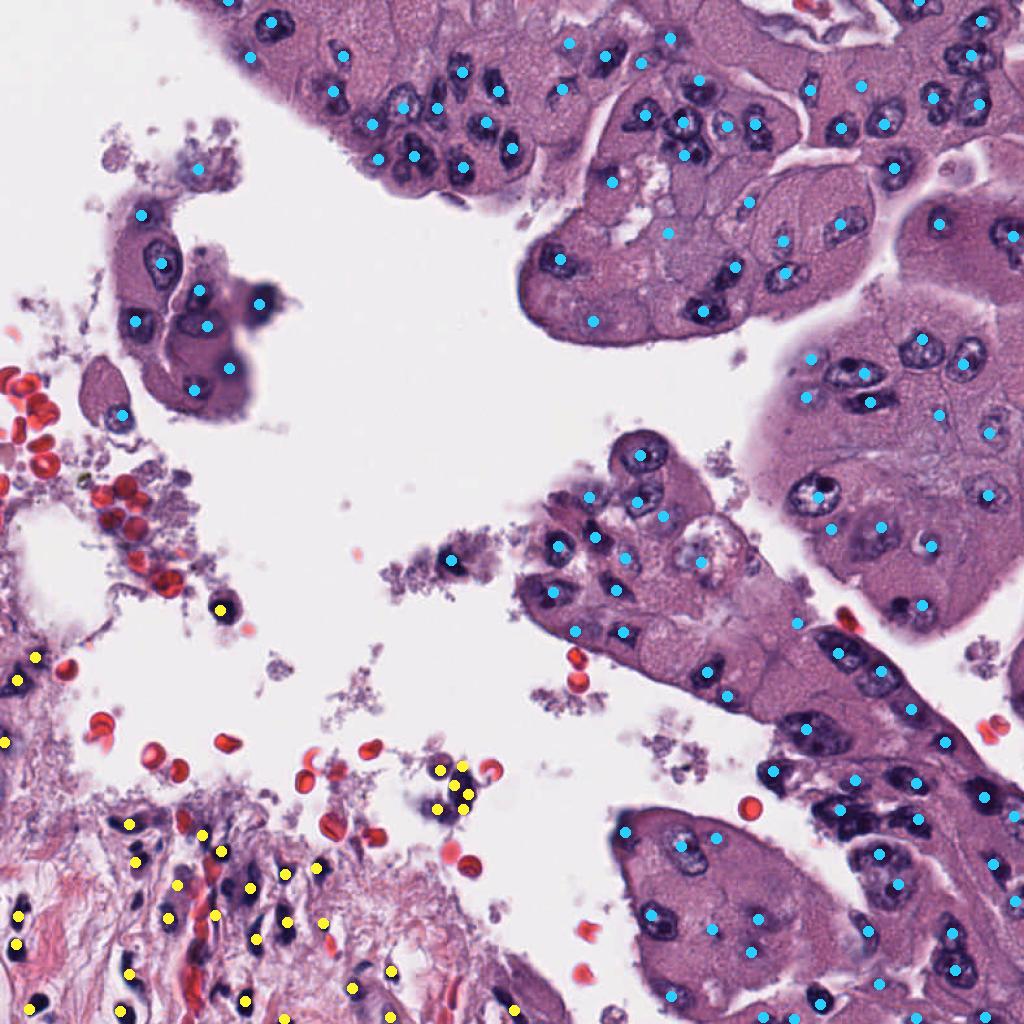}
        \caption{}
    \end{subfigure}
    \begin{subfigure}[t]{0.3\textwidth}
        \centering
        \includegraphics[width=0.9\textwidth]{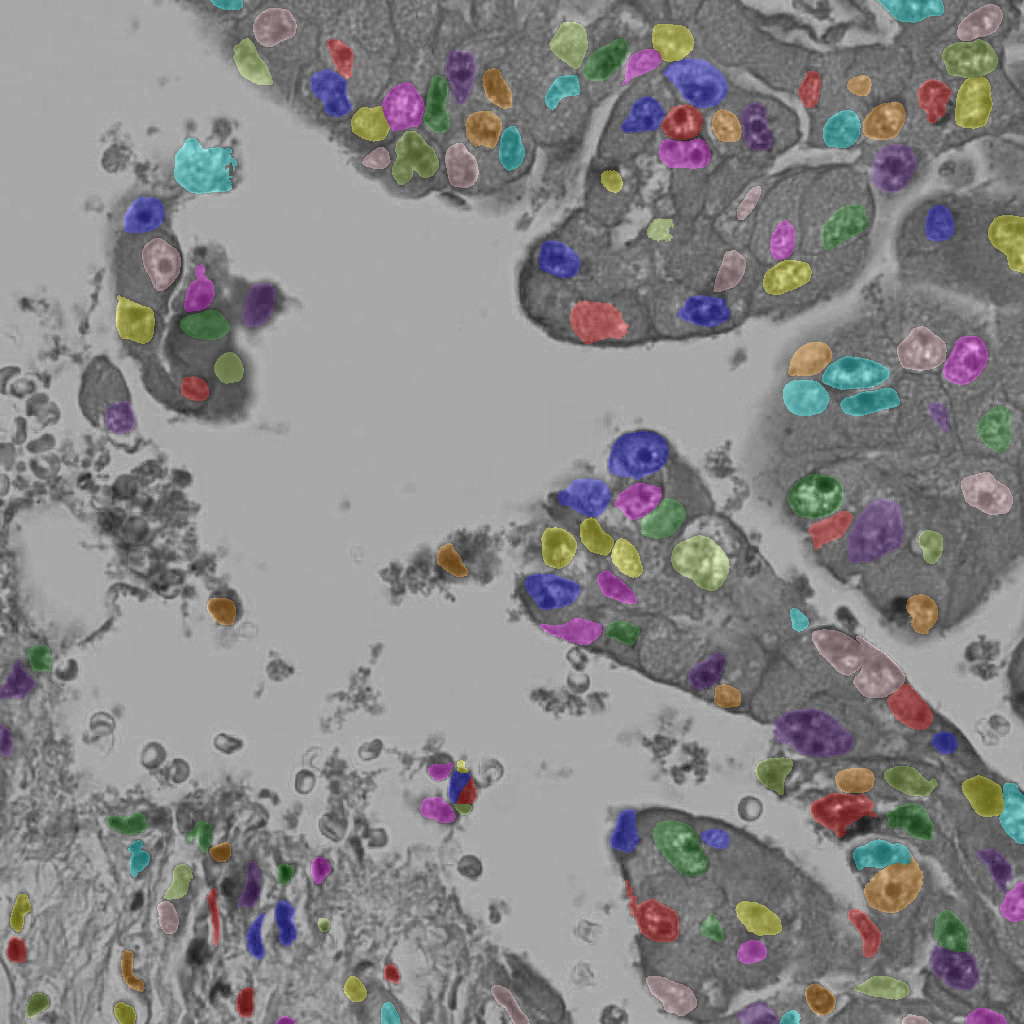}
        \caption{}
    \end{subfigure}
    \caption{(a) Original image, (b) Visualized point annotations, (c) Nuclick prediction}
    \label{fig:appx-nuclick}
\end{figure}

We utilize NuClick \cite{AlemiKoohbanani2020}, a pretrained nucleus, cell and gland segmentation model\footnote{Publicly available at \url{https://github.com/navidstuv/NuClick}, last accessed 24.11.2023}, to extend the cell annotations from centroid coordinates to segmentation maps, as visualized in Figure \ref{fig:appx-nuclick}. It relies on a guiding signal, in our case the cell point annotations, together with the input image for instance segmentation.
Pretrained weights are only available for nuclei segmentation\footnote{\url{https://drive.google.com/file/d/1MGjZs_-2Xo1W9NZqbq_5XLP-VbIo-ltA/view}, last accessed: 24.11.2023}, for this reason we extend the ground truth to a nuclei segmentation mask instead of a cell segmentation mask. 

The NuClick scripts required slight adaptation for our use case:
\begin{itemize}
    \item Read in cell points annotations in csv file format instead of mat.
    \item Read in images in jpg format instead of tif.
    \item Update tensorflow functions to avoid usage of deprecated functions.
\end{itemize}
The updated scripts are made public as a Github repository\footnote{NuClick repository with adaptations: \url{https://github.com/lely475/NuClick}}.

\section{Postprocessing for the hard IS ground truth format}
\label{appx:nuclick-postprocessing}
For the hard IS ground truth format, cell detection candidates are extracted by first combining the tumor and background cell prediction into a foreground prediction. The Otsu threshold is then applied on the foreground, resulting in a binary foreground map, from which holes and small objects are removed using the \textit{skimage.morphology.remove\_small\_objects}  and \textit{scipy.ndimage.binary\_fill\_holes} functions. Next, we calculate the Euclidean Distance Transform (EDT) on the binary image, revealing the distance of each pixel to the nearest background pixel and, consequently, highlighting potential cell instances as peaks. To identify these peaks, we employ \textit{skimage.feature.peak\_local\_max} function, assigning each peak a unique identifier. These identified points serve as markers in a marker-controlled watershed technique, facilitating the separation of connected objects within the binary foreground map. The inverse EDT (-EDT) is used as a cell border indicator. The final cell candidates are determined by locating the center of mass within each segmented instance of the watershed segmentation. The cell class is assigned as the class with the majority of class pixels in each cell instance.
\begin{figure}[htbp]
    \centering
    \includegraphics[width=\textwidth]{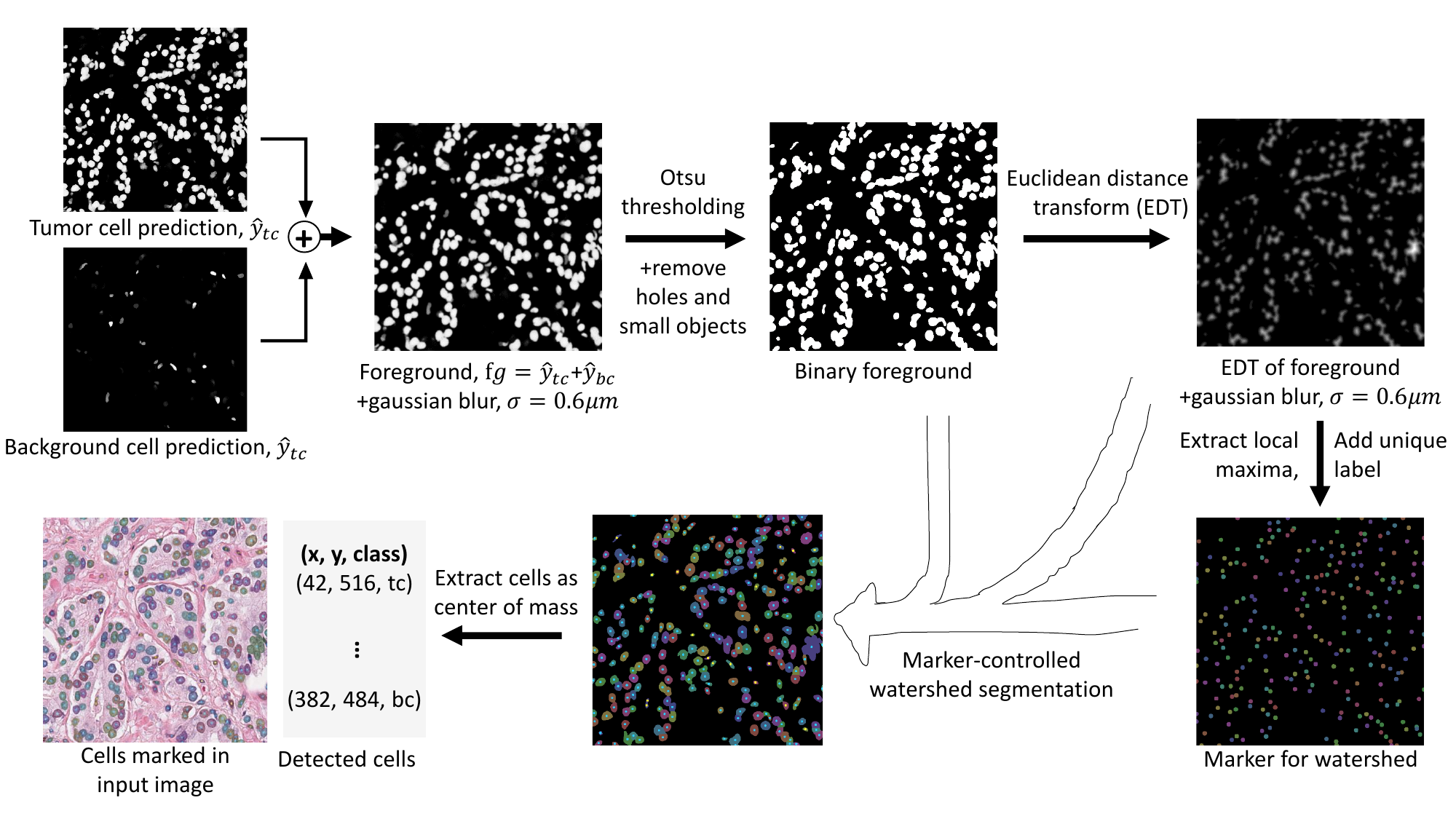}
    \caption{Postprocessing for hard IS ground truth format, tc: tumor cell class, bc: background cell class} %The input to the model is the image, or the image combined with the tissue segmentation prediction for the }
    \label{fig:appx-postprocessing-nuclick}
\end{figure}

\newpage
\section{Ablation study: Different $\sigma$ values for soft IS ground truth}
\label{appdx:sigma}
\begin{figure}[htbp]
    \centering
    \begin{subfigure}[t]{0.24\textwidth}
        \centering
        \includegraphics[width=\textwidth]{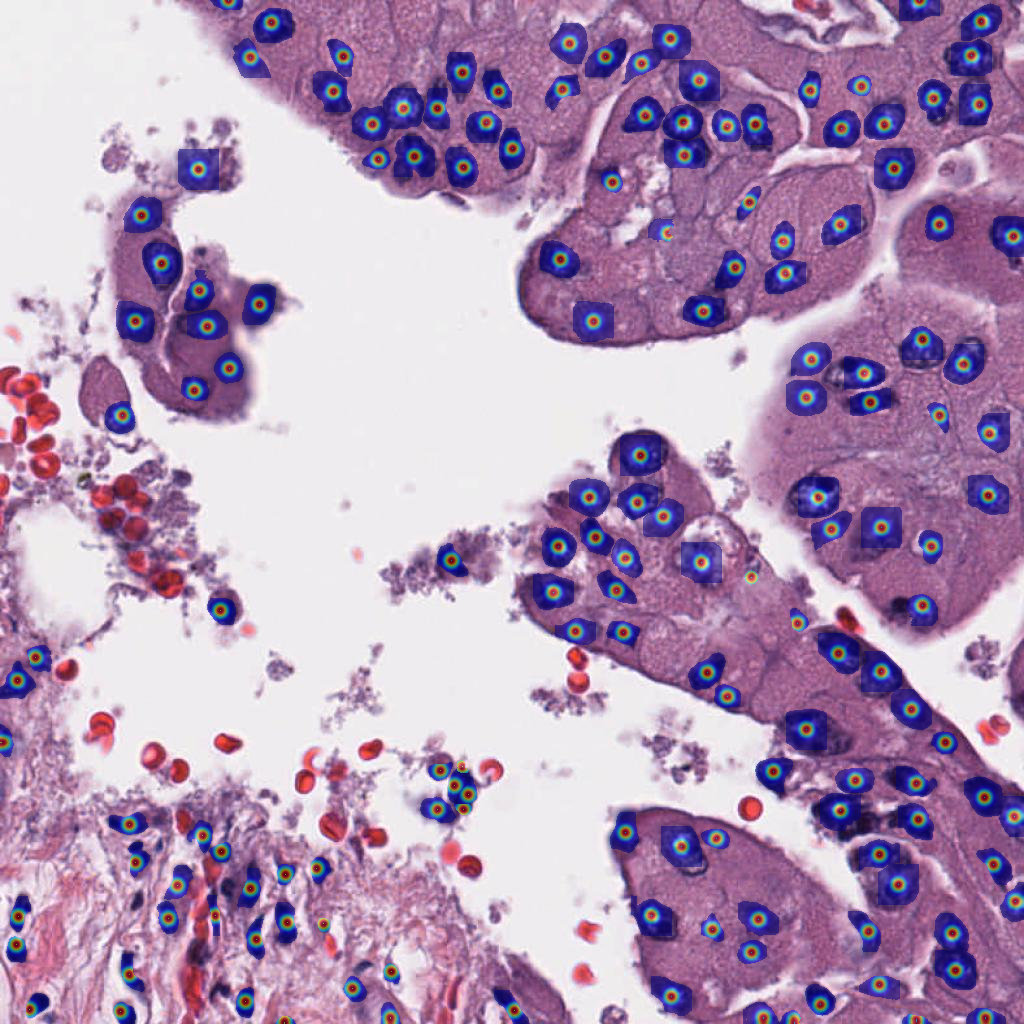}
        \caption{$\sigma=1\mu m$}
    \end{subfigure}
    \begin{subfigure}[t]{0.24\textwidth}
        \centering
        \includegraphics[width=\textwidth]{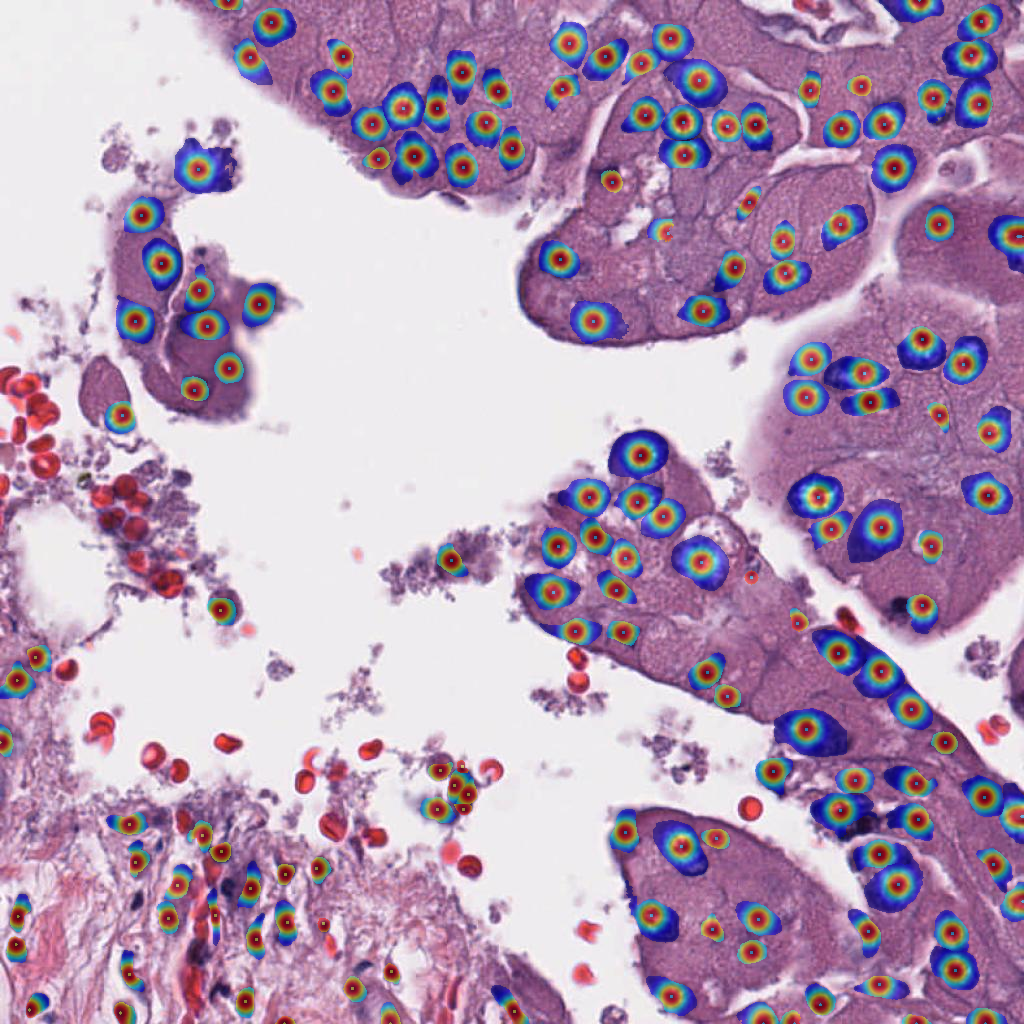}
        \caption{$\sigma=2\mu m$}
    \end{subfigure}
    \begin{subfigure}[t]{0.24\textwidth}
        \centering
        \includegraphics[width=\textwidth]{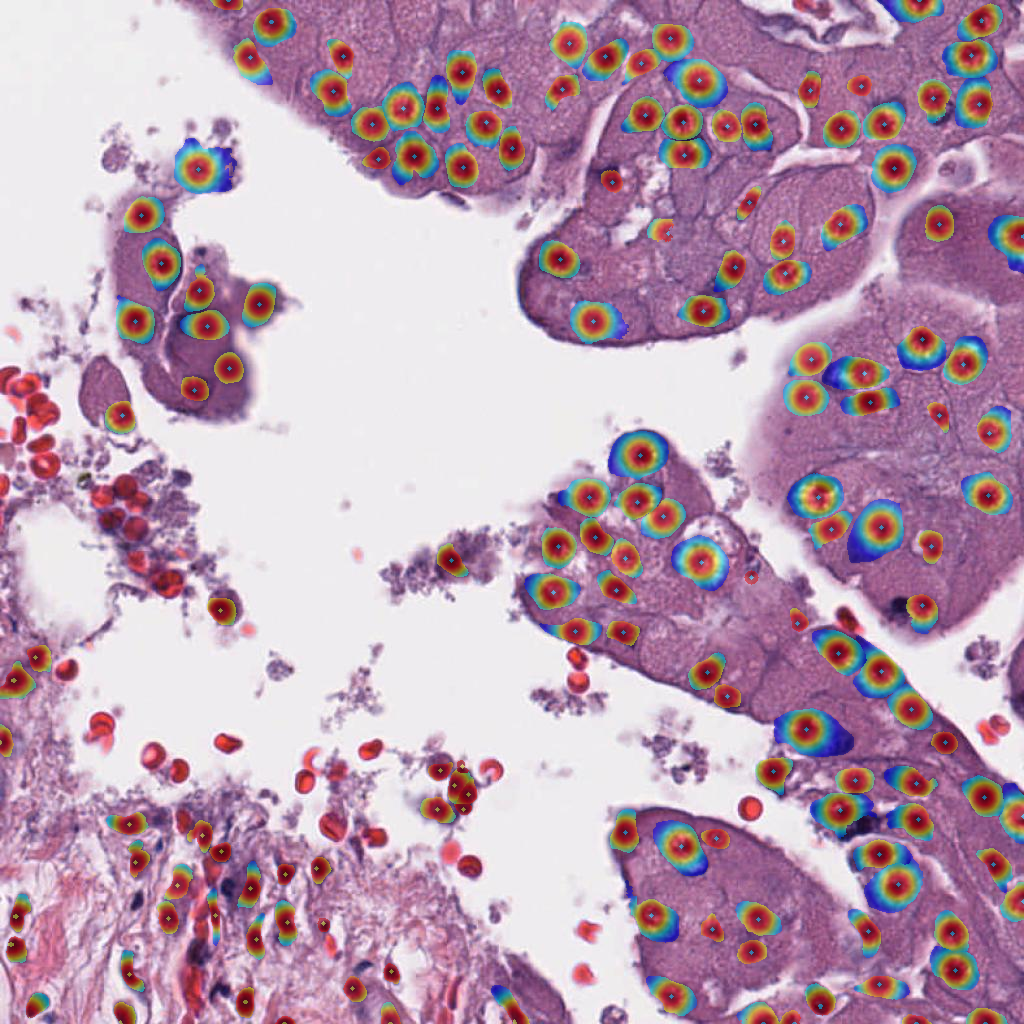}
        \caption{$\sigma=3\mu m$}
    \end{subfigure}
    \begin{subfigure}[t]{0.24\textwidth}
        \centering
        \includegraphics[width=\textwidth]{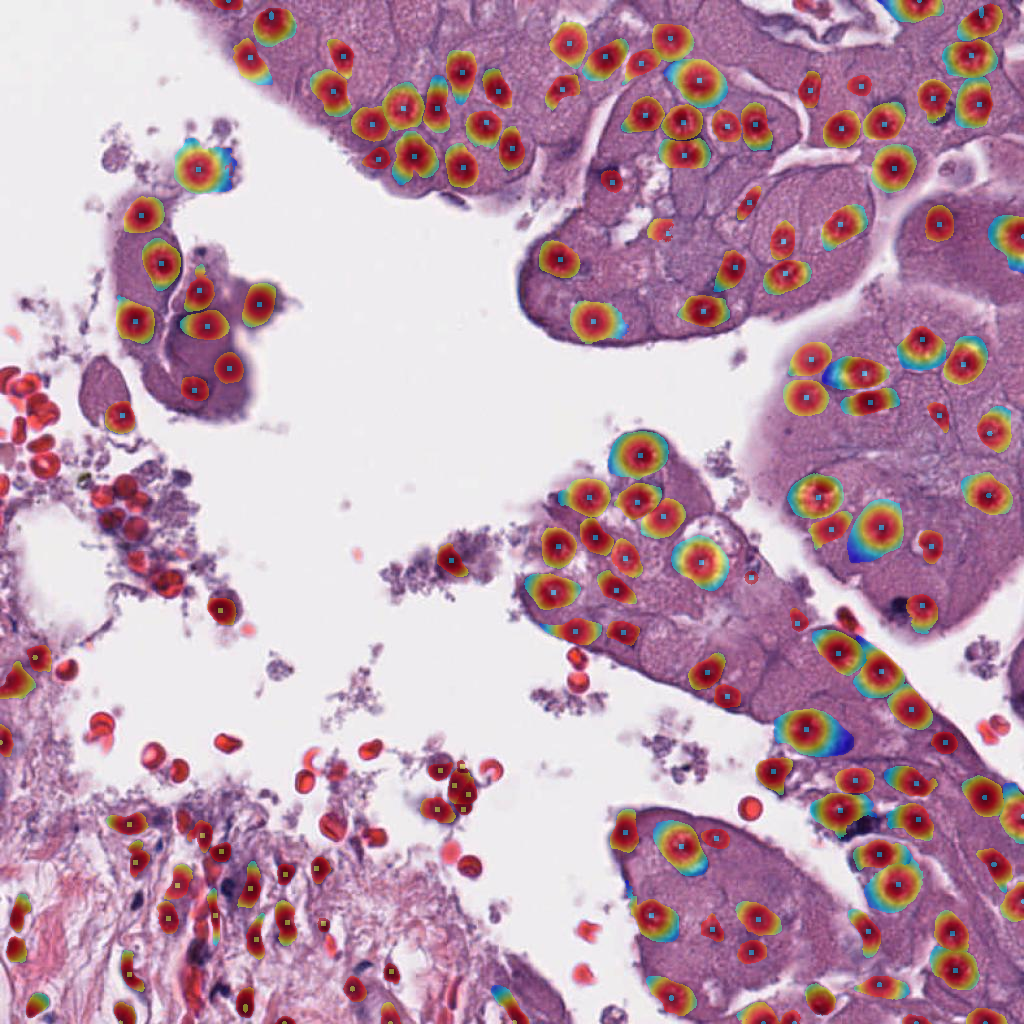}
        \caption{$\sigma=4\mu m$}
    \end{subfigure}
    \caption{Example of soft IS mask for $\sigma=[1,2,3,4]\mu m$}
    \label{fig:appx-sigma}
\end{figure}

The $\sigma=3\mu m$ value for soft IS was originally chosen based on a visual review. To investigate, whether there would be a more suitable $\sigma$ value, we conducted an ablation study with $\sigma=[1,2,3,4]\mu m$, analyzing the performance of the soft IS model. Figure \ref{fig:appx-sigma} shows an example of the soft IS mask for all considered $\sigma$ values. The internal train and validation set were used for training and evaluation.
The highest mean F1-Score is achieved for $\sigma=2\mu m$, while our initial choice $\sigma=3\mu m$ shows a slightly lower performance by -0.6\%.
Notably, the choice of $\sigma$ appears to be a compromise between precision and recall (Table \ref{tab:appx-sigma}). While lower $\sigma$ lead to a more precise cell detection, this comes at the cost of a higher number of missed cells (false negatives). The opposite is the case for larger $\sigma$ values. Overall, while the difference in F1-Score for $\sigma=2\mu m$ and $\sigma=3\mu m$ is minor, utilizing $\sigma=2\mu m$ for future work is recommended. 
\begin{table}[htb]
 \caption{Performance comparison of different $\sigma=[1,2,3,4]\mu m$ for the soft IS probability map on the internal validation set. We report the mean and std of 3 runs. %BC=Background Cell, TC=Tumor Cell
 }
  \centering
    \begin{tabular}{l|C{3cm}C{3cm}C{3cm}} 
     & F1-Score & Precision & Recall\\
    \midrule
   $\sigma=1\mu m$ & .5939$\pm0.0044$ & \textbf{.7695$\pm0.0092$} & .4760$\pm0.0035$\\
   $\sigma=2\mu m$ & \textbf{.6709$\pm0.0013$} & .7234$\pm0.0088$ & .6266$\pm0.0083$\\
   $\sigma=3\mu m$ & .6649$\pm0.0026$  & .6894$\pm0.0048$ & .6424$\pm0.0053$\\
   $\sigma=4\mu m$ & .6587$\pm0.0031$  & .6618$\pm0.0008$ & \textbf{.6562$\pm0.0056$}\\
    \bottomrule
  \end{tabular}
  \label{tab:appx-sigma}
\end{table}

\newpage
\section{Organ-wise performance on OCELOT validation set}
\label{appdx:organ-validation}
Table \ref{tab:validation-organ} and \ref{tab:validation-organ-tumor-sgm} detail the results for each approach on the OCELOT validation set.
Notably, while there is a trend for the soft IS, SoftCTM and TTA to improve performance for the majority of organs, there are also multiple organs for which one or multiple of these trends are not present. %In particular for kidney the highest mean F1 score is achieved for the hard IS and adding the tissue prediction reduced performance by 5\%. 
This highlights the challenge of training a model which can generalize well to different organs.
\begin{table}[htb]
 \caption{Comparison of per-organ mean F1 scores on the OCELOT validation set for cell detection. The CTM trained with the soft IS ground truth is denoted as SoftCTM. The highest score for the three ground truth formats is \underline{underlined}, the overall highest score is marked in \textbf{bold}.}
  \centering
  \begin{tabular}{lccccccc}
    \midrule
    & all & kidney & endometrium & bladder & prostate & stomach & head-neck \\
    & n=137 & n=41 & n=29 & n=29 & n=17 & n=12 & n=9 \\
     \midrule
    Circle &  .6781 & .6580 & .7167 & .5824 & \underline{.6174} & .6148 & .5470\\
   Hard IS & .6826 & \underline{\textbf{.7101}} & .7094  &  \underline{.5852} & .6076 & .6984 & .5401\\
   \textbf{Soft IS} & \underline{.6937} & .6934 & \underline{\textbf{.7437}}  &.5818 & .6136 & \underline{.7041} & \underline{.5682} \\
   \hline
    SoftCTM & .6875 & .6432 & .7258  & .6363 & .6328 & .7333 & .5832 \\
    SoftCTM + TTA & \textbf{.7046} & .6861 & .7397 & \textbf{.6364} & \textbf{.6385} & \textbf{.7412} & \textbf{.5848} \\
   % \hline
   %  TLLM & .7529 & .7200 & .7580  & .7200 & .7447 & .5716 & .7408 \\
   %  TLLM + TTA & .7562 & .7228 & .7622 & .7251  & .7464 & .5679 & .7444 \\
    \bottomrule
  \end{tabular}
  \label{tab:validation-organ}
\end{table}
\begin{table}[htb]
 \caption{Comparison of per-organ mean F1 scores on the OCELOT validation set for tissue segmentation. The tissue segmentation model is denoted as TSM. The highest score for each organ is marked in \textbf{bold}.}
  \centering
  \begin{tabular}{lccccccc}
    \midrule
    & all & kidney & endometrium & bladder & prostate & stomach & head-neck \\
    & n=130 & n=41 & n=25 & n=26 & n=16 & n=12 & n=10 \\
     \midrule
    TSM & .8511 & .8103 & .9258 & \textbf{.8270}  & .7914 & .6187  & \textbf{.7752} \\
    TSM + TTA & \textbf{.8571} & \textbf{.8281} & \textbf{.9307}  & .8186 & \textbf{.7970} &\textbf{.6243} & .7693 \\
    \bottomrule
  \end{tabular}
  \label{tab:validation-organ-tumor-sgm}
\end{table}

\end{document}